\documentclass[a4paper,fleqn]{cas-sc}

\usepackage[authoryear,longnamesfirst]{natbib}
\usepackage{wrapfig}
\usepackage[ruled,vlined]{algorithm2e}
\usepackage{algpseudocode} 
\usepackage{subfig}
\usepackage{multirow}
\graphicspath{ {Figures/} }
\def\tsc#1{\csdef{#1}{\textsc{\lowercase{#1}}\xspace}}
\tsc{WGM}
\tsc{QE}
\tsc{EP}
\tsc{PMS}
\tsc{BEC}
\tsc{DE}
\begin{document}
\let\WriteBookmarks\relax
\def\floatpagepagefraction{1}
\def\textpagefraction{.001}
\shorttitle{Impact of $h$-index on authors ranking}
\shortauthors{P. Khurana and K.Sharma}
%\begin{frontmatter}

\title [mode = title]{Impact of $h$-index on authors ranking: An improvement to the $h$-index for lower-ranked authors}                      
\tnotemark[1,2]

\author[1]{Parul Khurana}%[orcid=0000-0002-1690-7972]
%\ead{parul11183@gmail.com }
\credit{Collected the data and performed the analysis; Prepared figures and wrote the manuscript}

\author[2]{Kiran Sharma \corref{cor1}}[orcid=0000-0002-3797-7363]
\ead{kiran.sharma@bmu.edu.in}
\credit{Designed and performed the analysis; Prepared figures and wrote the manuscript}

\address[1]{School of Computer Applications, Lovely Professional University, Phagwara, Punjab-144401, India}

\address[2]{School of Engineering and Technology, BML Munjal University, Gurugram, Haryana-122413, India}
%\cortext[cor2]{Principal author}
\cortext[cor1]{Corresponding author}
%%%%%

%%%%%%%%%%%%%%%%%%%%%%%%%%%%%%
\begin{abstract}
In academia, the research performance of a faculty is either evaluated by the number of publications or the number of citations. Most of the time $h$-index is widely used during the hiring process or the faculty performance evaluation. The calculation of the $h$-index is shown in various databases; however, there is no systematic evidence about the differences between them. Here we analyze the publication records of 385 authors from Monash University (Australia) to investigate (i) the impact of different databases like Scopus and WoS on the ranking of authors within a discipline, and (ii) to complement the $h$-index, named $h_c$, by adding the weight of the highest cited paper to the $h$-index of the authors.  The results show the positive impact of $h_c$ on the lower-ranked authors in every discipline. Also, Scopus provides overall better ranking than WoS; however, for disciplines,  the ranking varies among Scopus and WoS.

\end{abstract}

%====================================
\begin{highlights}
\item To check the impact of $h$-index on authors ranking within a university based on Scopus and WoS.

\item The introduction of $h_c$ which is a complement to $h$-index, takes into account the weight of the highest cited paper.

\item Giving weight to highest cited paper significantly improves the performance of the lower-ranked authors, especially for $h\leq 10$. For higher rank, $h_c==h$ and for lower rank, $h_c \geq h$.

\item The performance of Scopus and WoS varies among disciplines, hence the ranking varies.

\end{highlights}

%====================================
\begin{keywords}
$h$-index \sep $g$-index \sep $h_c$-index \sep Authors ranking
\end{keywords}

\maketitle

%====================================
\section{Introduction}
Scientific evaluation is best carried out with the number of publications,  the number of citations, and contribution of an author to scientific knowledge and society \citep{martin1996use, garfield2006citation}. However, citation analysis plays a crucial role while evaluating the research performance of an individual in the academic community, that is why it acts as a key tool in scientometrics \citep{cronin1997comparative,  bornmann2005does, molinari2008new, bornmann2017measuring}. Along with the citations, the number of publications and $h$-index also have a strong stand in research evaluation~\citep{hirsch2005index}.  To perform the fair evaluation of an individual within a university/institution, funding bodies, scientific society, etc., it is the essential requirement that the considered scientometric parameters should be field, discipline, and time normalized \citep{waltman2016review}. However, with the rapid increase in the number of scholarly databases or libraries like Google Scholar, Scopus, Web of Science, Dimension, PubMed, etc., the choice of database consideration has become tedious due to the choice of considered journals \citep{bakkalbasi2006three, falagas2008comparison, meho2007impact,  mongeon2016journal,  martin2018google}.

As proposed by \cite{hirsch2005index} ``\textit{A scientist has index $h$ if $h$ of his/her $N_p$ papers have at least $h$ citations each and the other ($N_p -h$) papers have $\leq h$ citations each}.'' In bibliometric, $h$-index is considered as one of the important, robust, primitive, quantified, and a single measure used to evaluate the individual's work quality, impact, influence, and importance \citep{bar2007some, bar2008h}. However, the popularity and the interest gained by the $h$-index is due to its simple calculation among the scientific community  \citep{ball2005index,  dume2005high, glanzel2006h}. Instead of presenting individual values like the number of publications, and the number of citations, etc. which are giving a single dimension of the author's performance, $h$-index introduced multidimensional presentation (quantity and impact) and that is too with a single integer number \citep{maabreh2012survey}. Hence, it is considered as a balanced way to combine and evaluate the broad scientific impact of an author \citep{braun2005hirsch}. Gracza et al. have suggested $h$-index of an author as equivalent to the impact factor of a journal \citep{gracza2007impact}. Braun et al. have identified $h$-index as a measure of journal credibility and assessments as well \citep{braun2006hirsch}. Due to its popularity, different indexing databases like Scopus, WoS, etc. provides the calculated $h$-index of an author on their website \citep{egghe2008influence, abramo2010robust}.

Over the numerous advantages, it has some drawbacks too as mentioned by Hirsch in his core publication~\citep{hirsch2005index,  costas2007h}. To overcome such limitations, many new indices were proposed in this line and one such index is $g$-index~\citep{egghe2006theory,  batista2006possible, costas2007h, vinkler2007eminence, bornmann2008there, schreiber2008empirical, costas2008g}.
The limitations of not considering highly cited papers, have encouraged Leo Egghe to propose $g$-index in 2006 as follows: ``\textit{A set of papers has a g-index g if g is the highest rank such that the top $g$ papers have, together, at least $g^2$ citations. This also means that the top $g + 1$ papers have less than $(g + 1)^2$ papers}'' \citep{egghe2006improvement}. In comparison, the $g$-index shows improvement to $h$-index by giving more credit to highly cited papers and more discriminatory power to represent the scientific impact of author~\citep{schreiber2008influence, tol2008rational}.  Further, Leo Egghe introduced the concept of adding fictitious articles with 0 citations to overcome the limitations and complete the calculation of $g$-index~\citep{egghe2006theory}.

Over time, researchers have shown the use and importance of the $h$-index while calculating the ranking of authors, universities, the impact of a journal, etc.~\citep{bornmann2009state, torres2009ranking, vieira2009comparison}. Dunaiski et al. have evaluated the bias and performance of the authors over a range of citations; however, no significant differences between the globalized and averaged variants based on citations were found~\citep{dunaiski2019globalised}. Different approaches have been used in literature to analyze the author’s ranking. Authors have also shown the use of page rank algorithm on the author co-citations network to get the respective ranking~\citep{ding2009pagerank, nykl2015author, dunaiski2016evaluating, dunaiski2018author}.  Usman et al. have shown in their research the analysis of various assessment parameters like $h$-index, citations, publications, authors per paper, $g$-index, hg-index~\citep{alonso2010hg}, R-index~\citep{jin2007r}, e-index~\citep{zhang2009index}, h'-index~\citep{zhang2013h}, w-index~\citep{zhang2009proposal} etc.  to evaluate the authors ranking \citep{usman2020ranking}.

$h$-index ignores the highly cited papers that means it under-estimates the academic performance of the scientists. On the other hand, the $g$-index takes care of highly cited papers; however,  it reduces the impact of the highest cited paper significantly~\citep{ding2020exploring}. The present study resolves this problem by introducing a complementary index, named $h_c$,  that is complimenting $h$-index by including the weight of the highest cited paper while keeping the most important advantage of the $h$-index (a single-number to measure the academic performance).
In general, the aim of the study is to highlight the impact of the $h$-index on the author’s ranking while measuring their performance within an institution. Here we have examined the research contribution of an individual of Monash University in terms of the $h$-index and the weight of the highest cited paper from Scopus and WoS. This study aims to investigate the following points:

\begin{enumerate}
\item Impact of $h$-index on authors ranking based on different databases like Scopus and WoS.
\item Improving $h$-index by considering the weight of the highest cited paper that in turn improves the performance of the lower-ranked authors.
\end{enumerate}

The study is organized as follows: Section~\ref{sec:data} is on data description including data selection, data filtration, and discipline-wise publications analysis.
Results are explained in Section~\ref{sec:result}. Finally, the discussion and conclusion is presented in Section~\ref{sec:conclusion}.

%%====================================
%%====================================
\section{Data description}
\label{sec:data}
\subsection{Data selection}

%%%============= Figure-1: ============================
%%~~~~~~~~~~~~~~~%~~~~~~~~~~~~~~~%~~~~~~~~~~~~~~~
\begin{figure}[!h]
    \centering
    \includegraphics[width=0.85\linewidth]{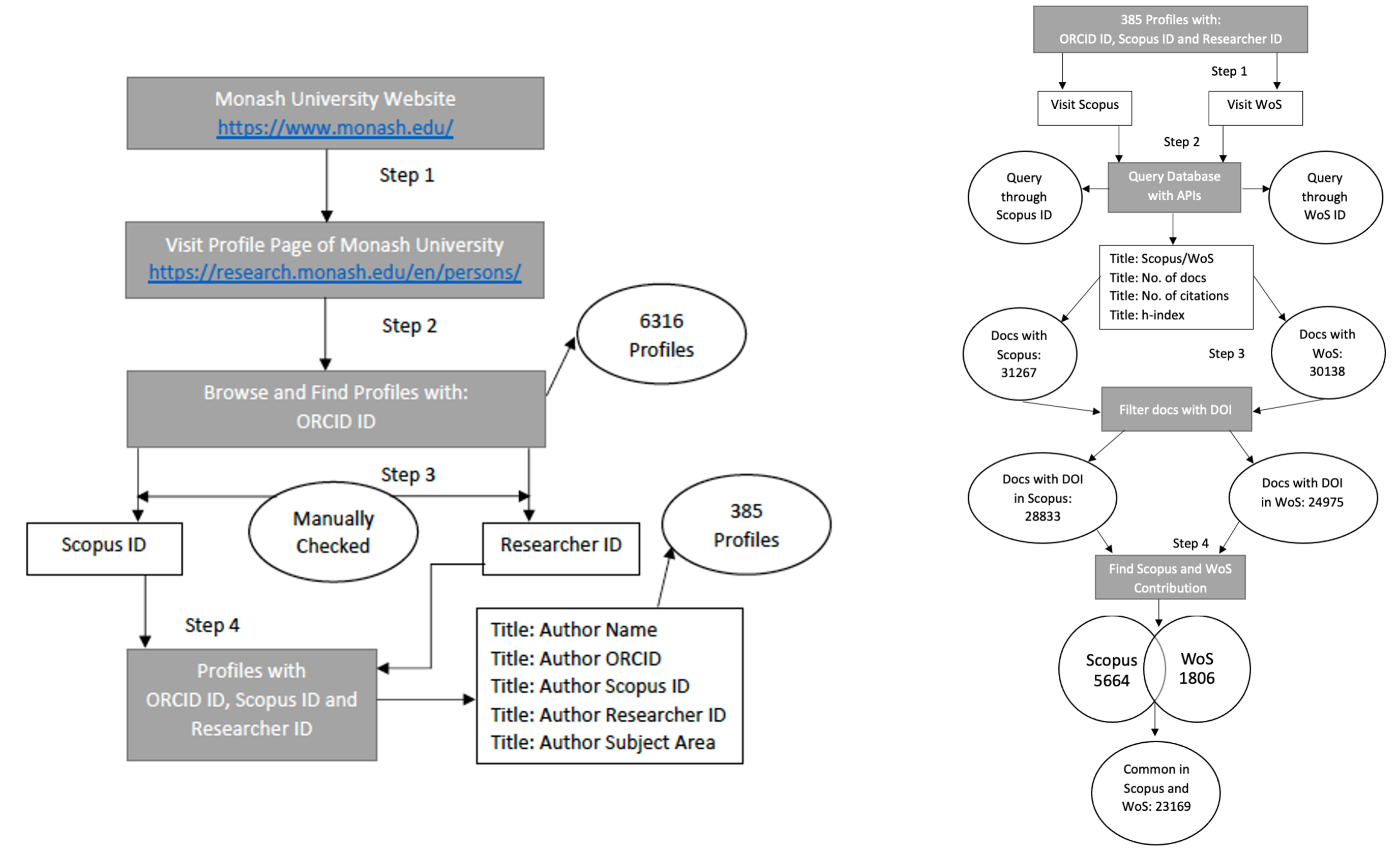}
    \caption{(left panel) Flowchart describing the process of visiting the author's profile.(right panel) Data extraction and filtration from Scopus and WoS.}
    \label{fig:1}
\end{figure}
%~~~~~~~~~~~~~~~%~~~~~~~~~~~~~~~%~~~~~~~~~~~~~~~
Data selection has two important aspects, first, what research question are we going to answer? and second, what is the required approach to answering the question? The goal is to study the ranking of authors based on the number of publications, citations, and $h$-index computed from Scopus and WoS.  The choice of the databases is arbitrary and is on the availability of the data. The very first challenge is the selection of the authors. On what basis an author should be selected is a major concern. To get detailed information about the scholarly data of any author we need authentic information. Indexing databases like Scopus or WoS track the author identity information with Author ID or Researcher ID or Orcid ID. There are two ways to approach this: one is to look for any open-source dataset like Kaggle, etc. and the other is to look manually at the university websites. Finally, we have found that Monash University, a public research university in Australia has provided the profiles of 6316 persons associated with the university at different designations and in different disciplines. There are three associated benefits with this dataset:
\begin{itemize}
\item First, all the profiles are sorted on the basis of the last name of persons.
\item Second, authors Orcid, Researcher, and Scopus IDs are mentioned.
\item Third, a search tab is given on the website to filter the profiles with at least 5 years or 10 years’ work with the university.
\end{itemize}

Then, we started searching the profiles of persons manually by opening all the profiles one by one. A further requirement of the research question is to identify those profiles which have all of three IDs: Orcid ID, Researcher ID, and Author ID. Orcid ID is a digital identifier and is used to uniquely identify authors across different platforms. Researcher ID is a digital identifier used by WoS to maintain the database of authors. Author ID is used by Scopus for the unique identification of authors. After checking each and every profile manually on the website of Monash University, we have considered the profile carrying all three IDs (see flowchart in Figure ~\ref{fig:1} (left panel)). To critically evaluate the identified research question, we have recorded the discipline of persons along with all three IDs (Orcid, Researcher, and Author). In the end, a sample of 385 persons from various disciplines is finalized and used to analyze the identified research questions.

%%===============================================
\subsection{Data filtration}
In order to perform the analysis, we have visited the profiles of 385 authors listed on the Monash University website \url{https://research.monash.edu/en/persons/}, which is freely available. Extracting different IDs (Orcid, Researcher, and Author) to avoid author ambiguity at any stage, later on, is the initial step of scrapping the data. So, the information extracted from Monash University regarding 385 authors is (see Fig.~\ref{fig:1} (left panel)):
\begin{itemize}
\item Author name
\item Authors Orcid, Researcher, and Scopus ID
\item Authors disciplines or subject categories
\end{itemize}
Further, we have extracted the following information from both Scopus and WoS using respective API’s (see Fig.~\ref{fig:1} (right panel):
\begin{itemize}
\item Author name, affiliation, $h$-index
\item Detailed records of the number of publications and citations received on those publications for all 385 authors
\item Doi’s of all publications
\end{itemize}
A total of 31267 documents are downloaded for all authors from Scopus and 30138 from WoS. To maintain the uniqueness among downloaded data, we have considered all the records with doi numbers only. Thus, we have filtered the number of documents in Scopus with doi as 28833 (92.2\%) and 24975 (82.9\%) in WoS. Further, we have filtered the common (both in Scopus and WoS) and unique (either Scopus or WoS) documents across both indexing databases. We have found that Scopus has 5664 (19.6\%) unique documents, and WoS has 1806 (7.2\%) unique documents. 23169 (80.4\% Scopus, 92.8\% WoS) of documents are common in both indexing databases.

We have categorized 385 records into five disciplines: \textit{Biochemistry and Molecular Biology}, \textit{Engineering},  \textit{Health and Medical Sciences}, \textit{Natural Sciences}, and \textit{Social Sciences}. In broader sense,  \textit{Biochemistry and Molecular Biology} includes microbiology, blood diseases, molecular sciences, etc. \textit{Engineering} includes civil, computer, electrical,  mechanical, etc.  \textit{Health and Medical Sciences} includes epidemiology,  medicine, physiology, etc.  \textit{Natural Sciences} includes physics, chemistry, mathematics, etc.  \textit{Social Sciences} includes sociology, economics,  psychology, etc.

%%============= Table-1: ============================
\begin{table}[!h]
\caption{A total number of authors,  total publications along with unique and common publications,  total citations, and publications received citations of five disciplines for both indexing databases (ID): Scopus (S) and WoS (W).}
\begin{tabular}{|l|c|c|c|c|c|c|c|}
\hline
Disciplines                                                                                   & \begin{tabular}[c]{@{}c@{}}Total\\ Authors\end{tabular} & ID & \begin{tabular}[c]{@{}c@{}}Total\\ Pubs\end{tabular} & \begin{tabular}[c]{@{}c@{}}Unique\\ Pubs\end{tabular} & \begin{tabular}[c]{@{}c@{}}Pubs\\ Received\\ Citations\end{tabular} & \begin{tabular}[c]{@{}c@{}}Total\\ Citations\end{tabular} & \begin{tabular}[c]{@{}c@{}}Common\\ Pubs\end{tabular} \\ \hline
\multirow{2}{*}{\begin{tabular}[c]{@{}c@{}}Biochemistry and Molecular Biology\end{tabular}} & \multirow{2}{*}{88}                                     & S  & 7095                                                 & 972                                                   & 6553                                                                & 294374                                                    & \multirow{2}{*}{5732}                                 \\ \cline{3-7}
                                                                                              &                                                         & W  & 6979                                                 & 366                                                   & 6241                                                                & 283221                                                    &                                                       \\ \hline
\multirow{2}{*}{Engineering}                                                                  & \multirow{2}{*}{60}                                     & S  & 5746                                                 & 1221                                                  & 4899                                                                & 137875                                                    & \multirow{2}{*}{3806}                                 \\ \cline{3-7}
                                                                                              &                                                         & W  & 5212                                                 & 270                                                   & 4418                                                                & 124490                                                    &                                                       \\ \hline
\multirow{2}{*}{\begin{tabular}[c]{@{}c@{}}Health and Medical Sciences\end{tabular}}        & \multirow{2}{*}{106}                                    & S  & 9687                                                 & 1786                                                  & 8503                                                                & 352286                                                    & \multirow{2}{*}{7125}                                 \\ \cline{3-7}
                                                                                              &                                                         & W  & 9807                                                 & 724                                                   & 7813                                                                & 330846                                                    &                                                       \\ \hline
\multirow{2}{*}{Natural Sciences}                                                             & \multirow{2}{*}{46}                                     & S  & 3266                                                 & 500                                                   & 2964                                                                & 101380                                                    & \multirow{2}{*}{2577}                                 \\ \cline{3-7}
                                                                                              &                                                         & W  & 3055                                                 & 123                                                   & 2773                                                                & 96889                                                     &                                                       \\ \hline
\multirow{2}{*}{Social Sciences}                                                              & \multirow{2}{*}{85}                                     & S  & 5473                                                 & 1185                                                  & 4762                                                                & 164524                                                    & \multirow{2}{*}{3929}                                 \\ \cline{3-7}
                                                                                              &                                                         & W  & 5085                                                 & 323                                                   & 4122                                                                & 141739                                                    &                                                       \\ \hline
\end{tabular}
\label{Table:pubs}
\end{table}
%%===============================================
\subsection{Discipline-wise publications analysis}
%%============= Figure-2: ============================
%~~~~~~~~~~~~~~~%~~~~~~~~~~~~~~~
\begin{figure}[!h]
    \centering
\includegraphics[width=0.9\linewidth]{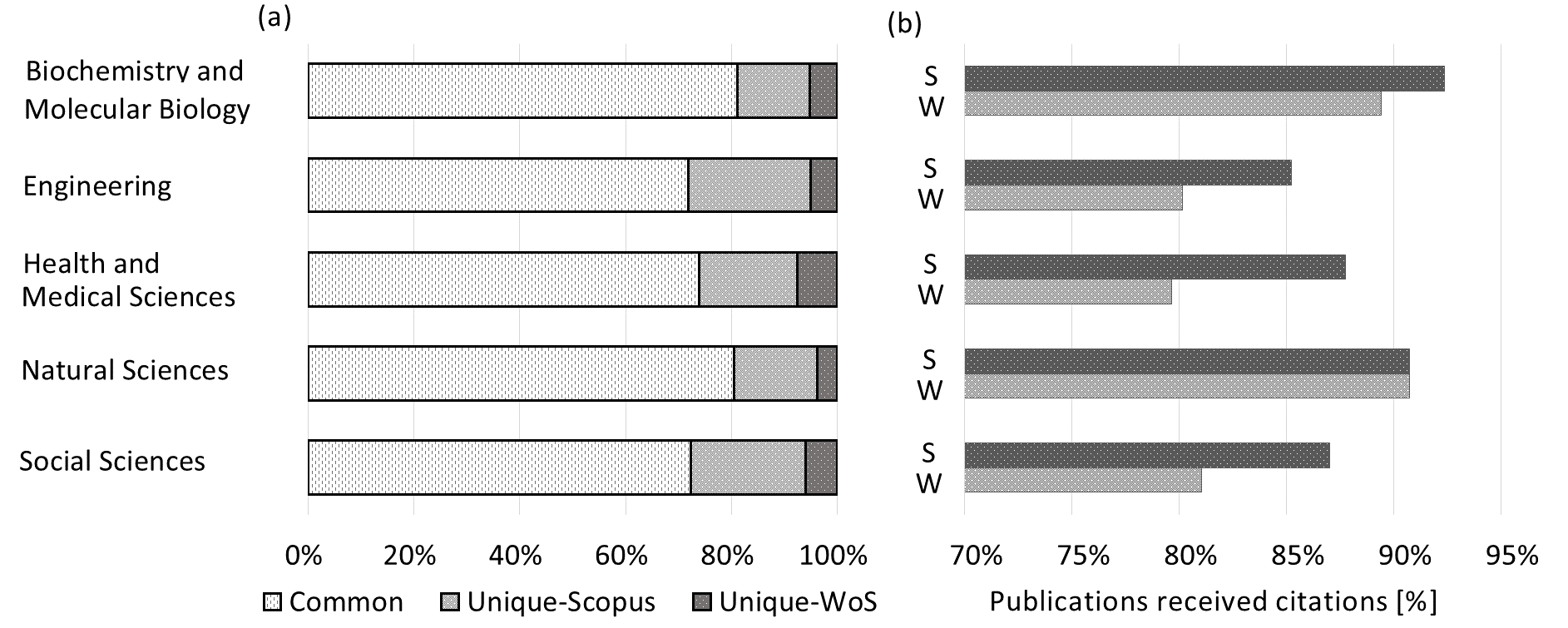} 
    \caption{(a) Discipline-wise proportion of the number of publications (\textit{common} and \textit{unique}) in Scopus and WoS. Unique publications are those which only appear either in Scopus (S) or in WoS (W). Common publications are the intersection of Scopus and WoS publications. (b) Number of publications received citations for Scopus and WoS.}
    \label{fig:2}
\end{figure}
%~~~~~~~~~~~~~~~~~~~~~~~~~~~~~~~~~~~~~~~~~~~~~~~~~~~
Figure~\ref{fig:2} (a) shows the proportion of number of common and unique publications in Scopus and WoS. \textit{Biochemistry and Molecular Biology} and \textit{Natural Sciences} contains more than 80\% of common publications whereas
\textit{Engineering},  \textit{Health and Medical Sciences}, and \textit{Social Sciences} contains more than 70\% of common publications.
Similarly, \textit{Engineering} and \textit{Social Sciences} contains more than 20\% of the unique Scopus publications, whereas WoS contains less than 10\% of unique publications.
On average, the number of common publications is 76\%, unique-Scopus is 19\%, and Unique-WoS is 5\%.
Hence,  the number of unique-Scopus publications is higher in all disciplines as compared to unique-WoS~\citep{falagas2008comparison, mongeon2016journal}. 

Figure~\ref{fig:2} (b) shows the proportion of publications that received citations from both Scopus and WoS. In \textit{Natural Sciences} both Socpus and WoS contains more than 90\% of publications that received citations. However, in \textit{Biochemistry and Molecular Biology} Scopus contains more than 90\% publications.  In Scopus, \textit{Engineering},  \textit{Health and Medical Sciences}, and \textit{Social Sciences} contains more documents that received citations as compared to WoS.
An index is being calculated on the number of publications and citations received on those publications. The variation in the number of publications in any database like Scopus, WoS, Google Scholar, MAG, PubMed, etc. affects the index computed on those publications~\citep{meho2007impact, martin2018google}. We will see this variation further in the study.  For the detail count of the number of authors,  publications, unique publications,  the number of publications that received citations, total citations, and common publications for both Scopus and WoS for five disciplines see~Table~\ref{Table:pubs}.

%%===============================================
\section{Results}
\label{sec:result}
\subsection{Authors ranking based on $h$, $g$, and $h_c$}

\begin{figure}[!h]
    \centering
    \includegraphics[width=0.75\linewidth]{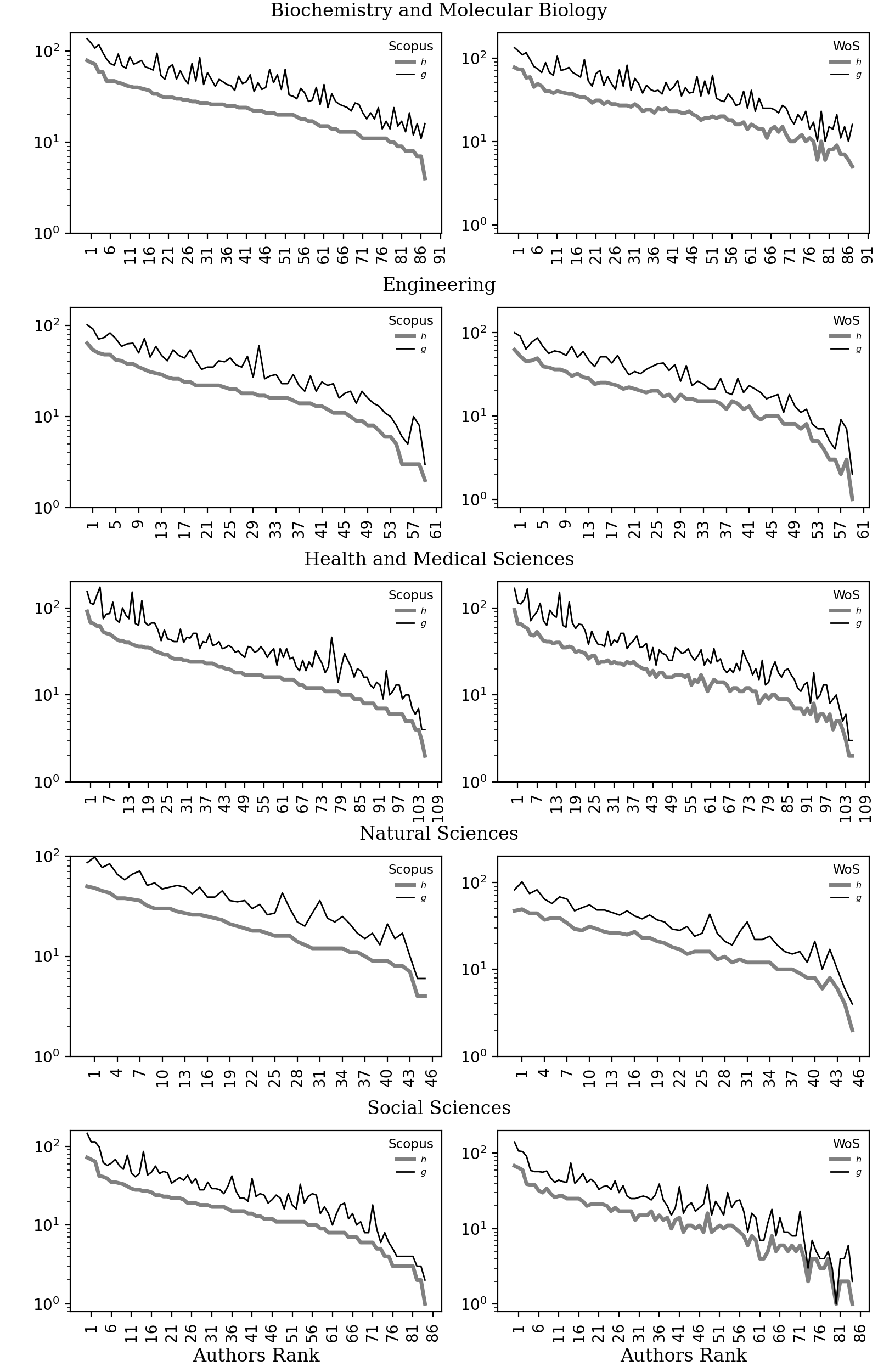}
     \llap{\parbox[b]{3.1in}{(a)\\\rule{0ex}{7.2in}}}
     \llap{\parbox[b]{0.75in}{(f)\\\rule{0ex}{7.2in}}}
     \llap{\parbox[b]{3.2in}{(b)\\\rule{0ex}{5.7in}}}
     \llap{\parbox[b]{0.85in}{(g)\\\rule{0ex}{5.7in}}}
     \llap{\parbox[b]{3.3in}{(c)\\\rule{0ex}{4.15in}}}
     \llap{\parbox[b]{0.95in}{(h)\\\rule{0ex}{4.15in}}}
     \llap{\parbox[b]{3.4in}{(d)\\\rule{0ex}{2.65in}}}
     \llap{\parbox[b]{1.0in}{(i)\\\rule{0ex}{2.6in}}}
     \llap{\parbox[b]{3.45in}{(e)\\\rule{0ex}{1.1in}}}
     \llap{\parbox[b]{1.1in}{(j)\\\rule{0ex}{1.1in}}}
     
    \caption{Ranking of 385 authors based on the $h$ and $g$ for both Scopus and WoS for five disciplines.  The ranks are sorted according to the $h$-index of Scopus.}
 \label{fig:3}
\end{figure}

Figure~\ref{fig:3} shows the ranking of authors for five disciplines in terms of $h$- and $g$- indices. The ranks are sorted according to Scopus $h$-index and $g$-index is being plotted for authors respectively.
A large number of fluctuation is observed for the authors with varying $h$-index in all disciplines.  In \textit{Biochemistry and Molecular Biology}, \textit{Health and Medical Sciences}, and \textit{Social Sciences}, the minimum value of $g$ is nearly double of $h$ in Scopus; whereas in WoS it is observed only in \textit{Biochemistry and Molecular Biology}.
Also, the average index value of $g$ is close to double of $h$ for all disciplines. 

%%============= Figure-3: ============================
%%~~~~~~~~~~~~~~~%~~~~~~~~~~~~~~~%~~~~~~~~~~~~~~~
%%~~~~~~~~~~~~~~~%~~~~~~~~~~~~~~~%~~~~~~~~~~~~~~~
%%============= Figure-4: ============================
%%~~~~~~~~~~~~~~~%~~~~~~~~~~~~~~~%~~~~~~~~~~~~~~~
\begin{figure}[!h]
    \centering
    \includegraphics[width=0.8\linewidth]{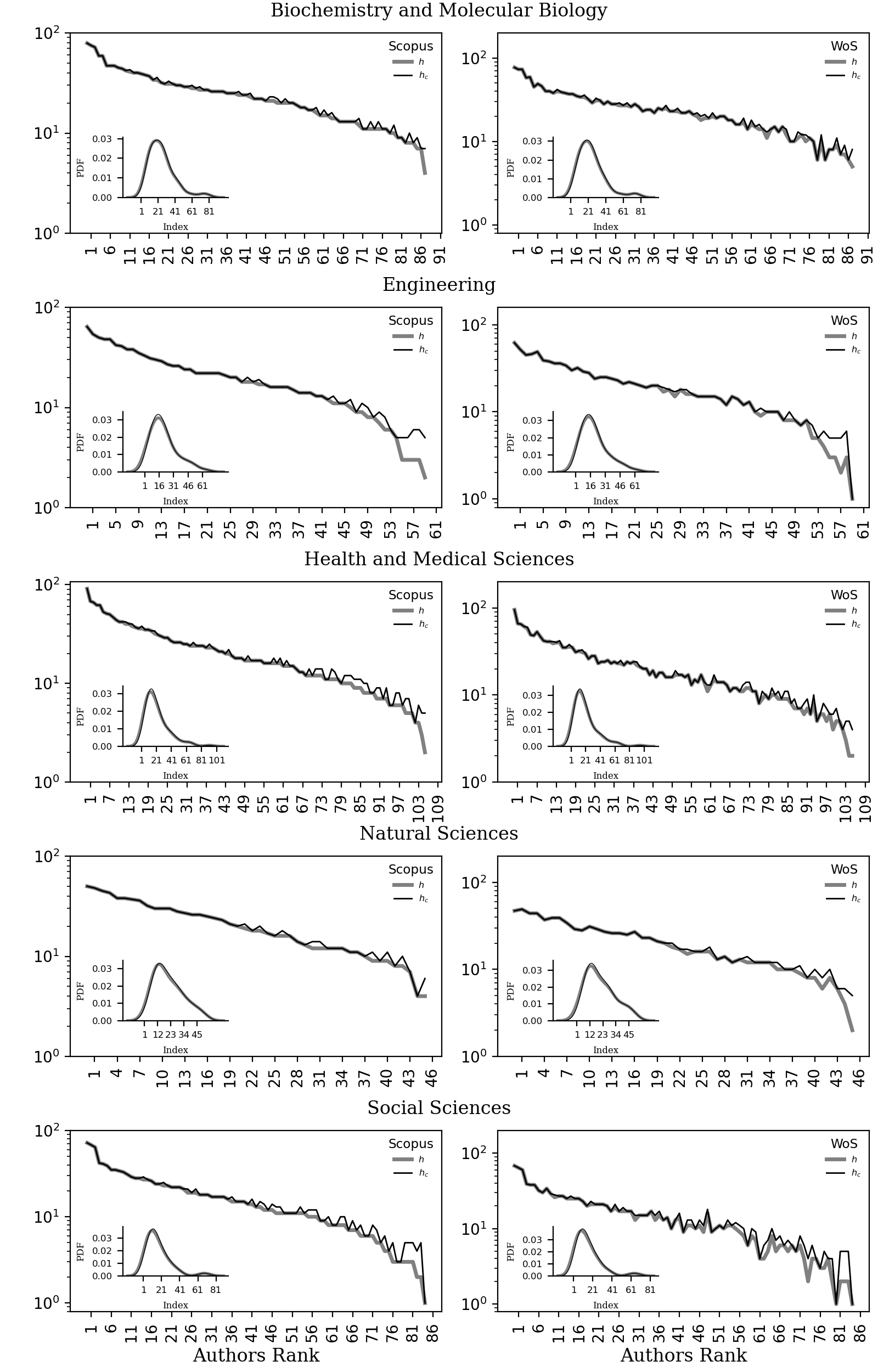}
          \llap{\parbox[b]{3.3in}{(a)\\\rule{0ex}{7.7in}}}
     \llap{\parbox[b]{0.8in}{(f)\\\rule{0ex}{7.7in}}}
            \llap{\parbox[b]{3.4in}{(b)\\\rule{0ex}{6.1in}}}
     \llap{\parbox[b]{0.9in}{(g)\\\rule{0ex}{6.1in}}}
            \llap{\parbox[b]{3.5in}{(c)\\\rule{0ex}{4.45in}}}
     \llap{\parbox[b]{1in}{(h)\\\rule{0ex}{4.45in}}}
            \llap{\parbox[b]{3.55in}{(d)\\\rule{0ex}{2.85in}}}
     \llap{\parbox[b]{1in}{(i)\\\rule{0ex}{2.85in}}}
            \llap{\parbox[b]{3.65in}{(e)\\\rule{0ex}{1.23in}}}
     \llap{\parbox[b]{1.1in}{(j)\\\rule{0ex}{1.23in}}}
    \caption{Ranking of 385 authors based on the $h$ and $h_c$ for both Scopus and WoS for five disciplines.  The ranks are sorted in descending order according to the $h$-index of Scopus. \textit{Inset}: shows the probability density function for both $h$ and $h_c$.  }
    \label{fig:4}
\end{figure}

%%~~~~~~~~~~~~~~~%~~~~~~~~~~~~~~~%~~~~~~~~~~~~~~~
\subsubsection{Definition of the $h_c$}
In this work, we have proposed a complementary analysis to the existing $h$-index, named $h_c$. Here, we are providing weight to the highest cited paper and add it to the $h$-index of an author. 
To get the weight of the highest cited paper, check the following condition
\begin{equation}
h^k < H_{cite},
\label{eq2}
\end{equation}
where $1 < h < H_{cite}$.  $H_{cite}$ is the count of highest cited paper of an author and $k$ is the weight of the highest cited paper and $k\geq 2$. Now $h_c$ is computed as
\begin{equation}
h_c = h+k.
\end{equation}
Here, $h_c \geq h$.For $k=1$ the condition~\ref{eq2} will always be \textit{True} and it will not add any value to the ranking. Hence, we reject the case.

\subsubsection{Comparative analysis of $h$ and $h_c$}
Table~\ref{Table:case_study} explains the three case studies representing different scenarios of the research productivity. $P_n$ represents the paper number, CR is citations received, $k$ is the calculated weight of the highest cited paper $H_{cite}$.

\begin{algorithm}[!h]
\SetAlgoLined
$i=2, k=0$\\
\While {$h^i <H_{cite}$}  {
        $k = i$\\
        $i = i+1$\\
}
$h_c = h+k$\\
\caption{Calculation of $h_c$}
\label{pseudo_code}
\end{algorithm}

\begin{itemize}
\item \textbf{Case I:} Here, $h = 4$ as 4 of the papers are having citations greater than or equal to 4 each.  $H_{cite} = 15$ for 10 publications (first paper of $P_n$ as papers are sorted in descending order according to citations received). 
From algorithm~\ref{pseudo_code}, we get $k=0$, hence, $h_c = h=4$. So, no change is observed.

\textbf{Case II:} Here, $h = 5$ and $H_{cite} = 65$. From algorithm~\ref{pseudo_code}, we get $k=2$, hence, $h_c = 7$.  So , it has raised the index value by 2.

\textbf{Case III:} Here, $h = 5$ and $H_{cite} = 205$.  From algorithm~\ref{pseudo_code}, we get $k=3$, hence, $h_c = 8$.  So , it has raised the index value by 3.
\end{itemize}

Figure~\ref{fig:4} shows the changes in $h_c$ with respect to $h$ for both Scopus and WoS.  The inset shows probability density function of $h$ and $h_c$. The minor shift of the $h_c$ towards the right shows the impact of $h_c$ on lower-ranked authors.
The major impact is on \textit{Social Sciences} where the index value of 32.9\% persons in Scopus and 37.6\% in WoS has been increased by $k=2$. Similarly, for $k=3$, 2.4\% in Scopus and 5.9\% in WoS has been increased.
The second highest is \textit{Health and Medical Sciences} with 31\% in Scopus and 26.4\% in WoS for $k=2$. There is negligible impact for $k=3$.
In \textit{Biochemistry and Molecular Biology}, 22.7\%  in Scopus and 23.9\% in WoS for $k=2$ and for $k=3$ the impact is negligible.
In \textit{Engineering}, 16.7\%  in Scopus and 15\% in WoS for $k=2$ and for $k=3$ it is 5\% for both Scopus and WoS. In \textit{Natural Sciences}, 19.6\%  in Scopus and 21.7\% in WoS for $k=2$ and for $k=3$ it is 2.2\% only for WoS.
The overall impact is 24.6\% in Scopus and 24.9\% in WoS for $k=2$ and 2.1\% in Scopus and 3\% in WoS for $k=2$.  The overall impact is almost same among Scopus and WoS; however, variations are visible in disciplines. In \textit{Health and Medical Sciences} WoS provides stable ranking whereas in \textit{Natural Sciences} and \textit{Social Sciences} Scopus provides stable ranking. In \textit{Biochemistry and Molecular Biology}, and \textit{Engineering} the difference is negligible.
In all disciplines, the minimum index value has been either increased, or remained same, i.e.  $h_c\geq h$ as shown in Table~\ref{Table:Stats}.  In some cases, $h_c$ is showing gain in minimum index value as compare to $g$.  There is no change in the maximum index value, i.e. $h==h_c$ and a slight deviation in median values for both Scopus and WoS. The average index value is almost same among Scopus and WoS for all disciplines.
%%============= Case Study ============================
\begin{table}[!h]
\centering
\caption{Demonstration of $h$ and $h_c$.}
\begin{tabular}{|c|c|c|c|c|c|c|c|c|c|c|c|c|c|c|c|c|}
\cline{1-5} \cline{7-11} \cline{13-17}
\multicolumn{5}{|c|}{Case I}                     &  & \multicolumn{5}{c|}{Case II}                     &  & \multicolumn{5}{c|}{ Case III}                       \\ \cline{1-5} \cline{7-11} \cline{13-17} 

\multicolumn{2}{|c|}{\textit{h}}                                                  & \multicolumn{3}{c|}{$h_c$}                                                                                        &  & \multicolumn{2}{c|}{\textit{h}}                                                   & \multicolumn{3}{c|}{$h_c$}                                                                                        & \multicolumn{1}{l|}{} & \multicolumn{2}{c|}{\textit{h}}                                                    & \multicolumn{3}{c|}{$h_c$}                                                                                        \\ \cline{1-5} \cline{7-11} \cline{13-17} 
\begin{tabular}[c]{@{}c@{}} $P_n$\end{tabular}              & CR              & $k$             & $h^k$             & \begin{tabular}[c]{@{}c@{}}$h^k < H_{cite}$\end{tabular}             &  & \begin{tabular}[c]{@{}c@{}}$P_n$\end{tabular}              & CR              & $k$             & $h^k$              & \begin{tabular}[c]{@{}c@{}}$h^k < H_{cite}$\end{tabular}            & \multicolumn{1}{c|}{} & \begin{tabular}[c]{@{}c@{}} $P_n$\end{tabular}              & CR               & $k$             & $h^k$              & \begin{tabular}[c]{@{}c@{}}$h^k < H_{cite}$\end{tabular}            \\ \cline{1-5} \cline{7-11} \cline{13-17} 
1                                                               & 15              & 2             & 16             &  F                                                                             &  & 1                                                               & 65              & 2             & 25              & T                                                                            & \multicolumn{1}{c|}{} & 1                                                               & 205              & 2             & 25              & T                                                                            \\ \cline{1-5} \cline{7-11} \cline{13-17} 
2 & 13 & \multicolumn{3}{c|}{} &  & 2                                                               & 9              & 3             & 125             & F                                                                            & \multicolumn{1}{c|}{} & 2                                                               & 150              & 3             & 125             & T                                                                            \\ \cline{1-5} \cline{7-11} \cline{13-17} 
3                                                               & 10              & \multicolumn{3}{c|}{\multirow{9}{*}{\begin{tabular}[c]{@{}c@{}}$h_c$ = $h + k$ \\ $h_c$ = 4+0 \\ $h_c$ = 4\end{tabular}}} &  & 3                                                               & 8              & \multicolumn{3}{c|}{\multirow{9}{*}{\begin{tabular}[c]{@{}c@{}}$h_c$ = $h + k$ \\ $h_c$ = 5+2 \\ $h_c$ = 7\end{tabular}}} & \multirow{9}{*}{}     & 3                                                               & 85               & 4             & 625             & F                                                                            \\ \cline{1-2} \cline{7-8} \cline{13-17} 
4                                                               & 7               & \multicolumn{3}{c|}{}                                                                                          &  & 4                                                               & 7              & \multicolumn{3}{c|}{}                                                                                          &                       & 4                                                               & 40               & \multicolumn{3}{c|}{\multirow{8}{*}{\begin{tabular}[c]{@{}c@{}}$h_c$ = $h + k$ \\ $h_c$ = 5+3 \\ $h_c$ = 8\end{tabular}}} \\ \cline{1-2} \cline{7-8} \cline{13-14}
5                                                               & 3               & \multicolumn{3}{c|}{}                                                                                          &  & 5                                                               & 5              & \multicolumn{3}{c|}{}                                                                                          &                       & 5                                                               & 25               & \multicolumn{3}{c|}{}                                                                                          \\ \cline{1-2} \cline{7-8} \cline{13-14}
6                                                               & 2               & \multicolumn{3}{c|}{}                                                                                          &  & 6                                                               & 5               & \multicolumn{3}{c|}{}                                                                                          &                       & 6                                                               & 5                & \multicolumn{3}{c|}{}                                                                                          \\ \cline{1-2} \cline{7-8} \cline{13-14}
7                                                               & 1               & \multicolumn{3}{c|}{}                                                                                          &  & 7                                                               & 2               & \multicolumn{3}{c|}{}                                                                                          &                       & 7                                                               & 4                & \multicolumn{3}{c|}{}                                                                                          \\ \cline{1-2} \cline{7-8} \cline{13-14}
8                                                               & 1               & \multicolumn{3}{c|}{}                                                                                          &  & 8                                                               & 2               & \multicolumn{3}{c|}{}                                                                                          &                       & 8                                                               & 4                & \multicolumn{3}{c|}{}                                                                                          \\ \cline{1-2} \cline{7-8} \cline{13-14}
9                                                               & 1               & \multicolumn{3}{c|}{}                                                                                          &  & 9                                                               & 1               & \multicolumn{3}{c|}{}                                                                                          &                       & 9                                                               & 2                & \multicolumn{3}{c|}{}                                                                                          \\ \cline{1-2} \cline{7-8} \cline{13-14}
10                                                              & 0               & \multicolumn{3}{c|}{}                                                                                          &  & 10                                                              & 0               & \multicolumn{3}{c|}{}                                                                                          &                       & 10                                                              & 1                & \multicolumn{3}{c|}{}                                                                                          \\ \cline{1-2} \cline{7-8} \cline{13-14}
\multicolumn{2}{|c|}{\begin{tabular}[c]{@{}c@{}}$h$ = 4\\ $H_{cite}$ = 15\end{tabular}} & \multicolumn{3}{c|}{}                                                                                          &  & \multicolumn{2}{c|}{\begin{tabular}[c]{@{}c@{}}$h$ = 5 \\ $H_{cite}$ = 65\end{tabular}} & \multicolumn{3}{c|}{}                                                                                          &                       & \multicolumn{2}{c|}{\begin{tabular}[c]{@{}c@{}}$h$ = 5 \\ $H_{cite}$ = 205\end{tabular}} & \multicolumn{3}{c|}{}                                                                                          \\ \cline{1-5} \cline{7-11} \cline{13-17} 
\end{tabular}
\label{Table:case_study}
\end{table}
%%~~~~~~~~~~~~~~~%~~~~~~~~~~~~~~~%~~~~~~~~~~~~~~~
%%============= Table-Stats: ============================
\begin{table}[!h]
\centering
\caption{Statistics of $h$, $h_c$ and $g$ in terms of min, max, median, average, and standard deviation (SD) of both indexing databases (ID) Scopus and WoS for all five disciplines.}
\begin{tabular}{|l|c|c|c|c|c|c|c|c|c|c|c|c|c|c|c|c|}
\hline
\multirow{2}{*}{Disciplines}                                                                  & \multirow{2}{*}{ID} & \multicolumn{3}{c|}{Min} & \multicolumn{3}{c|}{Max} & \multicolumn{3}{c|}{Median} & \multicolumn{3}{c|}{Average} & \multicolumn{3}{c|}{SD}  \\ \cline{3-17} 
                                                                                              &                     & \textit{h}   & $h_c$  & g   & \textit{h}  & $h_c$  & g    & \textit{h}    & $h_c$   & g    & \textit{h}   & $h_c$    & g     & \textit{h} & $h_c$   & g    \\ \hline
\multirow{2}{*}{\begin{tabular}[c]{@{}l@{}}Biochemistry and\\ Molecular Biology\end{tabular}} & S                   & 4            & 7   & 11  & 79          & 79  & 137  & 22            & 23   & 43   & 25.2         & 25.6  & 47.5  & 15.3       & 15.1 & 27.1 \\ \cline{2-17} 
                                                                                              & W                   & 5            & 6   & 10  & 77          & 77  & 133  & 22            & 22   & 41   & 24.5         & 25.0  & 46.2  & 15.0       & 14.8 & 27.3 \\ \hline
\multirow{2}{*}{Engineering}                                                                  & S                   & 2            & 5   & 3   & 64          & 64  & 102  & 18            & 18   & 31   & 20.7         & 21.2  & 36.2  & 14.0       & 13.5 & 23.0 \\ \cline{2-17} 
                                                                                              & W                   & 1            & 1   & 2   & 62          & 62  & 99   & 16            & 18   & 30   & 19.6         & 20.0  & 34.4  & 13.5       & 13.2 & 22.8 \\ \hline
\multirow{2}{*}{\begin{tabular}[c]{@{}l@{}}Health and\\ Medical Sciences\end{tabular}}        & S                   & 2            & 4   & 4   & 91          & 91  & 173  & 17            & 17   & 33   & 21.6         & 22.3  & 41.9  & 16.1       & 15.8 & 33.9 \\ \cline{2-17} 
                                                                                              & W                   & 2            & 4   & 3   & 95          & 95  & 168  & 16            & 16   & 30   & 20.6         & 21.2  & 39.6  & 16.1       & 15.9 & 33.9 \\ \hline
\multirow{2}{*}{Natural Sciences}                                                             & S                   & 4            & 4   & 6   & 50          & 50  & 98   & 18            & 19   & 36   & 21.2         & 21.6  & 38.1  & 12.2       & 11.9 & 22.1 \\ \cline{2-17} 
                                                                                              & W                   & 2            & 5   & 4   & 49          & 49  & 101  & 17            & 18   & 33   & 20.6         & 21.1  & 36.8  & 12.3       & 11.9 & 22.0 \\ \hline
\multirow{2}{*}{Social Sciences}                                                              & S                   & 1            & 1   & 2   & 72          & 72  & 146  & 13            & 14   & 25   & 17.0         & 17.7  & 31.6  & 13.9       & 13.6 & 26.7 \\ \cline{2-17} 
                                                                                              & W                   & 1            & 1   & 1   & 68          & 68  & 141  & 11            & 13   & 23   & 15.4         & 16.3  & 28.8  & 13.2       & 12.9 & 24.9 \\ \hline
\end{tabular}
\label{Table:Stats}
\end{table}
%##############################################################
%%======================================================
%%~~~~~~~~~~~~~~~~~~~~~~~~~~~~~~~~~~~~~~~~~~~~~~~~~~~~~
%%======================================================
\subsection{Impact of indexing databases on authors ranking}

\begin{figure}[!h]
    \centering
      \includegraphics[width=0.9\linewidth]{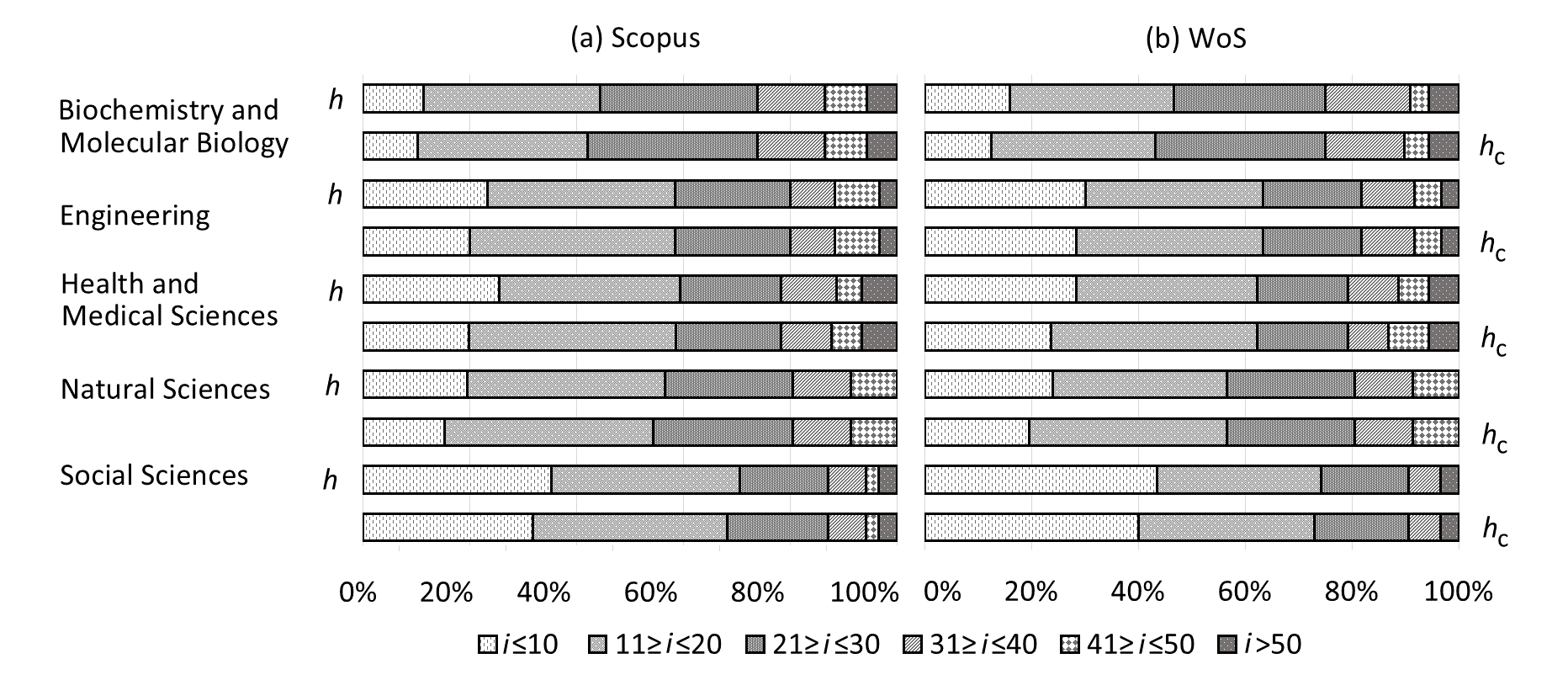} 
    \caption{Discipline-wise proportion of the number of authors with varying $h$ and $h_c$ in five disciplines for (a) Scopus and (b) WoS.   }
    \label{fig:5}
\end{figure}

Figure~\ref{fig:5} shows the distribution of 385 authors among five disciplines with varying $h$-index for (a) Scopus and (b) WoS.  We have placed the authors into six categories with varying $h$-index.  In total 23.4\% of authors are having $h \leq 10$ in Scopus and 28.6\% in WoS.  However, after complementing $h$ index as $h_c$,  this proportion lower down as 19.7\% for Scopus and 24.9\% for WoS for $h_c \leq 10$.
On the other hand, Scopus has 34.5\% and WoS has 32.2\% of authors in the range $11 \geq h \leq 20$, whereas this count has increased for $h_c$ as 36.6\% for Scopus and 34.8\% for WoS. Hence,  there is a 2\% gain in the authors count for both Scopus and WoS.
For higher-ranked authors ($h > 50$ ), Scopus has 4.4\% and WoS has 4.2\% of authors, and $h_c$ shows no impact on authors ranking at a higher level.
Table~\ref{Table:authors} shows the distribution of authors (in \%) based on $h$ and $h_c$ for both Scopus and WoS.
The proportion of authors having $h \leq 10$ is higher in WoS for all disciplines. In disciplines,\textit{Social Sciences} has highest count ( 35.3\%) and \textit{Biochemistry and Molecular Biology} has lowest count (11.4\%) for authors with $h \leq 10$ for both Scopus and WoS.  Further,  a significant change is noticed in \textit{Health and Medical Sciences}, and \textit{Natural Sciences} where the authors ranked $h \leq 10$  shows 5\% change in ranking from $h$ to $h_c$.  For rest of the disciplines the change is between 2-3\%.

Further, the shift in the author's ranking measured through the Spearman's rank correlation between $h$ and $h_c$ for both Scopus and WoS is given in Table~\ref{Table:corr}.  Spearman's rank correlation computes the association among two ranked variables. The initial rank of the author is $h$-index and the updated rank of the author is $h_c$ within a discipline for both Scopus and WoS. The major fluctuations are in the category of $h \leq 10$.  Both \textit{Biochemistry and Molecular Biology}, and  \textit{Natural Sciences} show large variation in Scopus and WoS. 
Major fluctuations are measured in \textit{Biochemistry and Molecular Biology},  \textit{Health and Medical Sciences},  and  \textit{Social Sciences}. Overall in \textit{Biochemistry and Molecular Biology} WoS performs better than Scopus and in \textit{Health and Medical Sciences} Scopus performs better than WoS. Hence, the performance of both Scopus and WoS varies among disciplines.

%%============= Figure-5: ============================
%~~~~~~~~~~~~~~~~~~~~~~~~~~~~~~~~~~~~~~~~~~~~~~~~~~~
%~~~~~~~~~~~~~~~%~~~~~~~~~~~~~~
%%============= Table-3 ============================

\begin{table}[!h]
\centering
\caption{Proportion of authors in five disciplines for varying $h$ and $h_c$ for Scopus (S) and WoS (W).}
\begin{tabular}{|c|c|c|c|c|c|c|c|c|c|c|c|c|c|}
\hline
\multirow{3}{*}{Disciplines}                                                                  & \multirow{3}{*}{ID} & \multicolumn{12}{c|}{No. of Authors [\%]}                                                                                                                                              \\ \cline{3-14} 
                                                                                              &                     & \multicolumn{2}{c|}{$h\leq10$} & \multicolumn{2}{c|}{$11 \geq h \leq20$} & \multicolumn{2}{c|}{$21 \geq h \leq 30$} & \multicolumn{2}{c|}{$31 \geq h \leq 40$} & \multicolumn{2}{c|}{$41 \geq h \leq 50$} & \multicolumn{2}{c|}{$h>50$} \\ \cline{3-14} 
                                                                                              &                     & \textit{h}  & \textit{$h_c$} & \textit{h}   & \textit{$h_c$}   & \textit{h}   & \textit{$h_c$}   & \textit{h}   & \textit{$h_c$}   & \textit{h}   & \textit{$h_c$}   & \textit{h}  & \textit{$h_c$} \\ \hline
\multirow{2}{*}{\begin{tabular}[c]{@{}c@{}}Biochemistry and\\ Molecular Biology\end{tabular}} & S                   & 11.4        & 10.2        & 33.0         & 31.8          & 29.5         & 31.8          & 12.5         & 12.5          & 8.0          & 8.0           & 5.7         & 5.7         \\ \cline{2-14} 
                                                                                              & W                   & 15.9        & 12.5        & 30.7         & 30.7          & 28.4         & 31.8          & 15.9         & 14.8          & 3.4          & 4.5           & 5.7         & 5.7         \\ \hline
\multirow{2}{*}{Engineering}                                                                  & S                   & 23.3        & 20.0        & 35.0         & 38.3          & 21.7         & 21.7          & 8.3          & 8.3           & 8.3          & 8.3           & 3.3         & 3.3         \\ \cline{2-14} 
                                                                                              & W                   & 30.0        & 28.3        & 33.3         & 35.0          & 18.3         & 18.3          & 10.0         & 10.0          & 5.0          & 5.0           & 3.3         & 3.3         \\ \hline
\multirow{2}{*}{\begin{tabular}[c]{@{}c@{}}Health and\\ Medical Sciences\end{tabular}}        & S                   & 25.5        & 19.8        & 34.0         & 38.7          & 18.9         & 19.8          & 10.4         & 9.4           & 4.7          & 5.7           & 6.6         & 6.6         \\ \cline{2-14} 
                                                                                              & W                   & 28.3        & 23.6        & 34.0         & 38.7          & 17.0         & 17.0          & 9.4          & 7.5           & 5.7          & 7.5           & 5.7         & 5.7         \\ \hline
\multirow{2}{*}{Natural Sciences}                                                             & S                   & 19.6        & 15.2        & 37.0         & 39.1          & 23.9         & 26.1          & 10.9         & 10.9          & 8.7          & 8.7           & 0.0         & 0.0         \\ \cline{2-14} 
                                                                                              & W                   & 23.9        & 19.6        & 32.6         & 37.0          & 23.9         & 23.9          & 10.9         & 10.9          & 8.7          & 8.7           & 0.0         & 0.0         \\ \hline
\multirow{2}{*}{Social Sciences}                                                              & S                   & 35.3        & 31.8        & 35.3         & 36.5          & 16.5         & 18.8          & 7.1          & 7.1           & 2.4          & 2.4           & 3.5         & 3.5         \\ \cline{2-14} 
                                                                                              & W                   & 43.5        & 40.0        & 30.6         & 32.9          & 16.5         & 17.6          & 5.9          & 5.9           & 0.0          & 0.0           & 3.5         & 3.5         \\ \hline
\end{tabular}
\label{Table:authors}
\end{table}

%%==============================================

%%============= Table-3 ============================

\begin{table}[!h]
\caption{Spearman's rank correlation shows the changes among the ranking.}
\begin{tabular}{|c|c|c|c|c|c|c|c|}
\hline
\multirow{2}{*}{Disciplines}                                                                  & \multirow{2}{*}{ID} & \multicolumn{6}{c|}{Spearman's rank correlation between $h$ and $h_c$}                                                                                  \\ \cline{3-8} 
                                                                                              &                     & \textit{h$\leq$10} & \textit{11$\geq$h$\leq$20} & \textit{21$\geq$h$\leq$30} & \textit{31$\geq$h$\leq$40} & \textit{41$\geq$h$\leq$50} & \textit{h$>$50} \\ \hline
\multirow{2}{*}{\begin{tabular}[c]{@{}l@{}}Biochemistry and\\ Molecular Biology\end{tabular}} & S                   & 0.77 & 0.96    & 0.96    & 0.97    & 0.96    & 1.00                       \\ \cline{2-8} 
                                                                                              & W                   & 0.85 & 0.96    & 0.95    & 0.98    & 1.00    & 1.00                       \\ \hline
\multirow{2}{*}{Engineering}                                                                  & S                   & 0.90 & 0.98    & 1.00    & 1.00    & 1.00    & 1.00                       \\ \cline{2-8} 
                                                                                              & W                   & 0.93 & 0.98    & 1.00    & 1.00    & 1.00    & 1.00                       \\ \hline
\multirow{2}{*}{\begin{tabular}[c]{@{}l@{}}Health and\\ Medical Sciences\end{tabular}}        & S                   & 0.92 & 0.93    & 0.94    & 0.97    & 1.00    & 1.00                       \\ \cline{2-8} 
                                                                                              & W                   & 0.91 & 0.96    & 0.86    & 0.96    & 1.00    & 1.00                       \\ \hline
\multirow{2}{*}{Natural Sciences}                                                             & S                   & 0.83 & 0.95    & 1.00    & 1.00    & 1.00    & -                          \\ \cline{2-8} 
                                                                                              & W                   & 0.87 & 0.92    & 1.00    & 1.00    & 1.00    & -                          \\ \hline
\multirow{2}{*}{Social Sciences}                                                              & S                   & 0.93 & 0.96    & 0.96    & 1.00    & 1.00    & 1.00                       \\ \cline{2-8} 
                                                                                              & W                   & 0.92 & 0.94    & 0.95    & 1.00    & -       & 1.00                       \\ \hline
\end{tabular}
\label{Table:corr}
\end{table}
%~~~~~~~~~~~~~~~~~~~~~~~~~~~~~~~~~~~~~~~~~~~~~~~~~~~

\begin{figure}%
    \centering
  \includegraphics[width=0.75\linewidth]{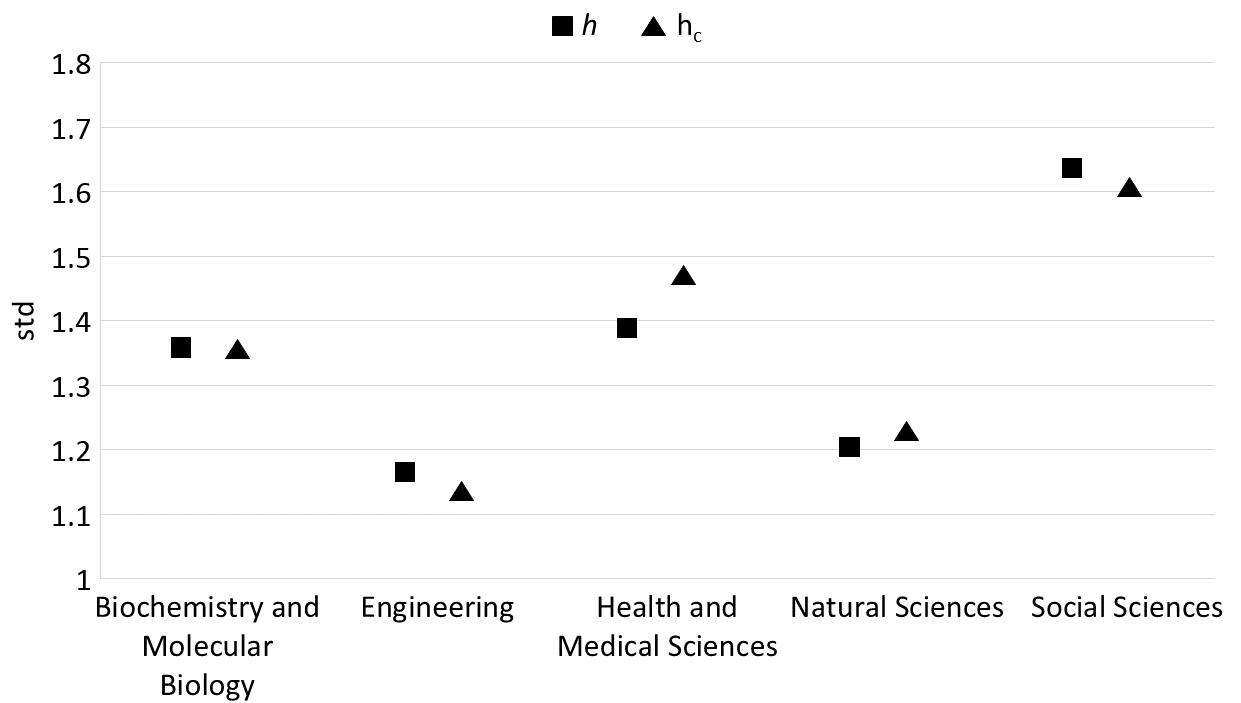} 
    \caption{Discipline-wise standard deviation of $h$ and $h_c$. The deviation is computed on the difference between Scopus $h$-index and WoS $h$- index (filled square) and Scopus $h_c$ and WoS $h_c$ (filled up-triangle). }
    \label{fig:6}
\end{figure}

%%============= Figure-6: ============================
Further, we have computed the difference between Scopus $h$-index and WoS $h$- index and calculated the standard deviation of the data for all five disciplines. Similarly, the standard deviation is calculated for $h_c$ and is shown in Figure~\ref{fig:6}.  The deviation in \textit{Biochemistry and Molecular Biology} and in \textit{Natural Sciences} is same.  The deviation is slightly lower for \textit{Engineering} and \textit{Social Sciences} whereas it is slightly higher for \textit{Health and Medical Sciences}. In general,  $h_c$ is not much deviated from $h$ and the results show a slight improvement to $h$ especially to lower-ranked profiles. 
%~~~~~~~~~~~~~~~~~~~~~~~~~~~~~~~~~~~~~~~~~~~~~~~~~~~
%~~~~~~~~~~~~~~~%~~~~~~~~~~~~~~

%%============= Figure-7: ============================
%~~~~~~~~~~~~~~~~~~~~~~~~~~~~~~~~~~~~~~~~~~~~~~~~~~~
%\begin{figure}[!h]
%    \centering
%  \includegraphics[width=0.8\linewidth]{Figure7.png} 
%       \llap{\parbox[b]{4.2in}{(a)\\\rule{0ex}{7.7in}}}
%     \llap{\parbox[b]{4.3in}{(b)\\\rule{0ex}{6.05in}}}
%          \llap{\parbox[b]{4.3in}{(c)\\\rule{0ex}{4.45in}}}
%    \llap{\parbox[b]{4.35in}{(d)\\\rule{0ex}{2.8in}}}
%     \llap{\parbox[b]{4.4in}{(e)\\\rule{0ex}{1.2in}}}
%         \llap{\parbox[b]{2.7in}{(f)\\\rule{0ex}{7.7in}}}
% 		\llap{\parbox[b]{2.78in}{(g)\\\rule{0ex}{6.05in}}}
%          \llap{\parbox[b]{2.8in}{(h)\\\rule{0ex}{4.45in}}}
%    \llap{\parbox[b]{2.85in}{(i)\\\rule{0ex}{2.8in}}}
%     \llap{\parbox[b]{2.9in}{(j)\\\rule{0ex}{1.2in}}}
%      \llap{\parbox[b]{1.25in}{(k)\\\rule{0ex}{7.7in}}}
%	\llap{\parbox[b]{1.25in}{(l)\\\rule{0ex}{6.05in}}}
%          \llap{\parbox[b]{1.35in}{(m)\\\rule{0ex}{4.45in}}}
%    \llap{\parbox[b]{1.4in}{(n)\\\rule{0ex}{2.8in}}}
%     \llap{\parbox[b]{1.45in}{(o)\\\rule{0ex}{1.2in}}}
%
%    \caption{Comparison among Scopus and WoS for all three indices: $h$, $g$ and $h_c$.  The $h_c$ index has significantly improved the ranking of lower ranked authors. The higher rank would always have maximum value of $h$-index.  }
%    \label{fig:7}
%\end{figure}
%%~~~~~~~~~~~~~~~%~~~~~~~~~~~~~~
%=========================
\section{Discussion and conclusion}
\label{sec:conclusion}
In this study, we have extended the pioneering work of J. E. Hirsch by simply focusing on one of the limitations of the $h$-index~\citep{costas2007h, egghe2010hirsch}. $h$-index focuses on both quantity and impact of publications but ignored the impact of highly cited papers which under-estimates the importance of the work.
After the introduction of the $h$ index,  many variants of $h$ have been proposed by scientists in order to improve the research evaluation of an individual~\citep{schreiber2010twenty}.  Some became meaningful like the $g$-index; however, each and every index is lacking in one or the other sense.  Based on the $h$-index we have introduced $h_c$ which is a complementary approach to the $h$-index. $h_c$ takes care of one of the limitations of the $h$-index, especially $h$ ignores the highly cited papers. $h_c$ works on $h$-index by adding weight to the highest cited paper of an individual. 
The main goals of the study are: (i) to overcome the limitation of $h$-index by taking into account the weight of the highest cited paper; (ii) an improvement to $h$-index which significantly improve the ranking of scientists having lower rank within a discipline; (iii) to analyze the impact of different databases on the authors ranking based on $h$-index;  and (iv) to analyze the impact of the databases on different disciplines which in turn affect the ranking of scientists.

Hirsch mentioned that the $h$-index cannot be compared among scientists of different disciplines.  In our study, we have worked on five disciplines to check how the ranking of authors based on the $h$-index varies among Scopus and WoS.  The new index $h_c$ is computed on the same ranking and we found that in total the major fluctuations appeared for the authors ranked $h\leq 10$ for both Scopus and WoS; however, this variation/fluctuation is higher in WoS as compared to Scopus.  In discipline-wise analysis, the results vary among Scopus and WoS. For \textit{Engineering}, \textit{Health and Medical Sciences}, and \textit{Natural Sciences}, WoS shows less variation as compared to Scopus. On the other hand, in \textit{Biochemistry and Molecular Biology}, and \textit{Social Sciences} Scopus has less variation as compared to WoS. Hence, in discipline-wise analysis WoS gives better ranking in some disciplines and in other Scopus does; however, the overall performance of Scopus is measured better.

The advantage of $h_c$ is that it is not sensitive to disciplines. Also, in some cases $h_c$ has performed better than $g$ too. Due to its simplicity and in complement to $h$-index, $h_c$ could provide a useful insight towards the young or lower-ranked authors which in turn significantly improves the ranking of an individual within a discipline. It could be helpful in an institution/ university for the internal ranking of the faculties within a discipline.  It also highlights the importance of the work carried out by an individual as it takes into account the $h$-index along with the contribution of the highest-cited paper.
The present work has some limitations too. The number of authors for the analysis of  \textit{Engineering} and \textit{Natural Sciences} disciplines is a bit low; however, we are able to get the essence of the ranking.  The larger sample can increase the universality of the results. For future study, the same concept can be extended towards the ranking of the organizations and journals.
%%=============================================
\section*{Acknowledgment}
Both Scopus and WoS data has been downloaded from the Northwestern University, USA.

\printcredits
\bibliographystyle{cas-model2-names}

\bibliography{cas-refs}

\end{document}